# Giant reduction of thermal conductivity in twinning superlattice InAsSb nanowires

Lorenzo Peri[1,†,¶], Domenic Prete[1,¶], Valeria Demontis[1], Valentina Zannier[1], Francesca Rossi[2], Lucia Sorba[1], Fabio Beltram[1] and Francesco Rossella[1,3,*]

[1] NEST, Scuola Normale Superiore and Istituto Nanoscienze-CNR, Pisa, Italy
[2] IMEM CNR
[3] Dipartimento di Scienze Fisiche, Informatiche e Matematiche, Università di Modena e Reggio Emilia, Modena, Italy
† present address: Cambridge, UK
¶ equally contributing author
* Corresponding author

**ABSTRACT**
Semiconductor nanostructures hold great promise for high-efficiency waste heat recovery exploiting thermoelectric energy conversion, a technological breakthrough that could significantly contribute to providing environmentally friendly energy sources as well as in enabling the realization of self-powered biomedical and wearable devices. A crucial requirement in this field is the reduction of the thermal conductivity of the thermoelectric material without detrimentally affecting its electrical transport properties.
In this work we demonstrate a drastic reduction of thermal conductivity in III-V semiconductor nanowires due to the presence of intentionally realized periodic crystal lattice twin planes. The electrical and thermal transport of these nanostructures, known as twinning superlattice nanowires, have been probed and compared with their twin-free counterparts, showing a one order of magnitude decrease of thermal conductivity while maintaining unaltered electrical transport properties, thus yielding a factor ten enhancement of the thermoelectric figure of merit, ZT. Our study reports for the first time the experimental measurement of electrical and thermal properties in twinning superlattice nanowires, which emerge as a novel class of nanomaterials for high efficiency thermoelectric energy harvesting.

**INTRODUCTION**
The urgent need for locally available and non-intermittent energy sources for the powering of wearable, implantable and biomedical devices has recently boosted theoretical and experimental studies on thermoelectric energy conversion [1], which entails the generation of an electrical current or voltage from a temperature gradient, leading to conversion of potentially wasted heat into usable electrical energy. A good thermoelectric material should possess high electrical conductivity, to facilitate the flow of charge carriers (e.g. electrons), while possessing low thermal conductivity, to minimize irreversible flow of heat [2]. In the context of the Onsager-De Groot-Callen theory [3], the performance at the temperature T of a thermoelectric material can be quantified with the thermoelectric figure of merit $ZT = \frac{\sigma S_e^2 T}{\kappa}$ , where σ and κ are the electrical and thermal conductivities, respectively, and $S_e$ is the electronic Seebeck coefficient of the material [4-5]. Striving to achieve higher figures of merit, research efforts have mainly focused on different ways to enhance the electrical conductivity. However, it is evident from the formula of ZT that another way to optimize ZT is to reduce the thermal conductivity - usually dominated by the lattice contribution in semiconductors at room temperature - without impacting the electrical transport properties.
In this frame, semiconductor nanowires (NWs) are extremely promising [6-11], as their high aspect ratio leads to an increase of phonon scattering at the lateral boundaries of the nanostructure. This effect, known as the Casimir effect, leads to a strong reduction of the effective phonon mean free path with respect to the intrinsic value of the bulk material [12], yielding to an effective reduction of the thermal conductivity [13, 14, 15]. For instance, InAs NWs - usually grown with a wurtzite (WZ) crystal structure - have been widely studied for thermoelectric applications [16, 17]. Unfortunately, despite the Casimir effect, single crystal nanowires made of common III-V or IV semiconductors still cannot compete with the industry standard thermoelectric materials (e.g. $Bi_2Te_3$) because their figure of merit is still significantly lower (between one and two orders of magnitude [18]) than the state of art thermoelectric materials employed in thermoelectric generators.

Improvement is still possible in III-V semiconducting nanowires by employing polytypic structures known as twin superlattices (TSLs) [19-24]. These structures involve the periodic repetition of twin planes, i.e., stacking faults inducing a rotation of the orientation of the crystal lattice with respect to the lattice portion preceding it, all along the NW growth direction [25]. The crystal structure around a twin plane can also be seen as an inclusion of a different polytype with respect to the dominant one, such as a wurtzite (WZ) segment in a zincblende (ZB) structure [26]. On the one hand, the random occurrence of this kind of stacking faults, often observed in vapor liquid solid (VLS)-grown III-V NWs, is usually considered as a defect, as they can have detrimental effects on the electrical properties of the material [27]. On the other hand, the intentional introduction of these planar defects in a periodic fashion to form a crystal phase superlattice [28] can strongly modify the electronic band structure of the material, e.g. promoting the formation of electronic minibands [29]. A controlled and periodic repetition of twin planes in TSL NWs can be achieved by finely tuning the growth parameters, the NW diameter and the introduction of controlled amount of impurities [30, 31].

In this work, we experimentally investigate $InAs_{1-x}Sb_x$ NWs as novel nanomaterials for the realization of next-generation thermoelectric energy converters. A small and controlled amount of Sb added to InAs NWs changes the crystal structure from pure WZ to pure ZB, via intermediate stacking-faults and periodic-twinning regimes [30, 32] and these nanostructures are excellent candidates to feature strongly reduced thermal conductivity [33-34] with unaffected electrical conductivity [29,35]. For the first time, we experimentally demonstrate that periodic twinning in semiconductor nanowires allows to engineer heat transport inside the material to achieve a drastic decrease of its thermal conductivity together with unaffected electrical conductivity, hence improving its thermoelectric properties. We have measured the electrical and thermal transport properties of TSL InAsSb NWs with different twin plane periodicity and compared them with those of pure ZB InAsSb NWs and pure WZ InAs NWs, revealing a strongly improved thermoelectric figure of merit, calculated by exploiting Mott's formula [36] for the estimation of the Seebeck coefficient. We believe that this work paves the way for the development of high-performance thermoelectric generators based on III-V semiconductor nanostructures, along with giving new life to the research on phonon engineering in semiconducting nanomaterials.

**RESULTS & DISCUSSION**
**Twinning Superlattice Nanowires**
The nanowire samples grown for the present study include WZ InAs NWs, ZB InAsSb NWs and TSL InAsSb NWs. High resolution TEM (HRTEM) images of the different InAsSb NWs are reported in **Figure 1**, while **Table 1** reports the growth conditions employed to achieve the different nanostructures. Figure 1 (a-d) shows the HRTEM images of the TSL NWs (samples A, B, C, D of Table 1). Their structural and morphological properties have been compared with pure ZB InAsSb NW (sample E; zoomed HRTEM in Figure 1(e)). We have found that the crystal structure, that is WZ in the InAs NW stem grown before the InAsSb segment, changes to ZB as soon as some Sb is incorporated, as already reported elsewhere [32, 37]. Simultaneously, there is a non-negligible lateral growth occurring along with the axial elongation during the InAsSb segment growth, so that the final NW diameter is larger than the nanoparticle (NP) size. NWs with Sb fraction higher than 7% (sample E in Table 1) have a pure ZB crystal structure without any structural defect, while samples with Sb fraction in the range between 1% and 6% (A-D in the table) show a periodic repetition of twin planes for the whole length, making them TSL NWs. Most importantly, we have found that the twin periodicity is affected by both chemical composition (Sb fraction) and diameter of the gold NP used for the VLS growth. In fact, for the same NP diameter, NWs with higher Sb incorporation feature longer periodicity. Fixed the Sb fraction, larger NPs give as well TSL NWs with longer twin periodicity (see for example samples D and A). This trend is consistent with what reported for different materials [26, 27] and explained with the distortion of the nanoparticle in response to the evolution of the hexagonal cross-section of NWs with non-parallel {111} side faces during the growth [26]. Phonon scattering occurring at the twin planes (pictorially represented in Figure 1(f)-bottom right panel) is expected to drastically reduce the thermal conductivity while preserving the electrical conduction at room temperature.

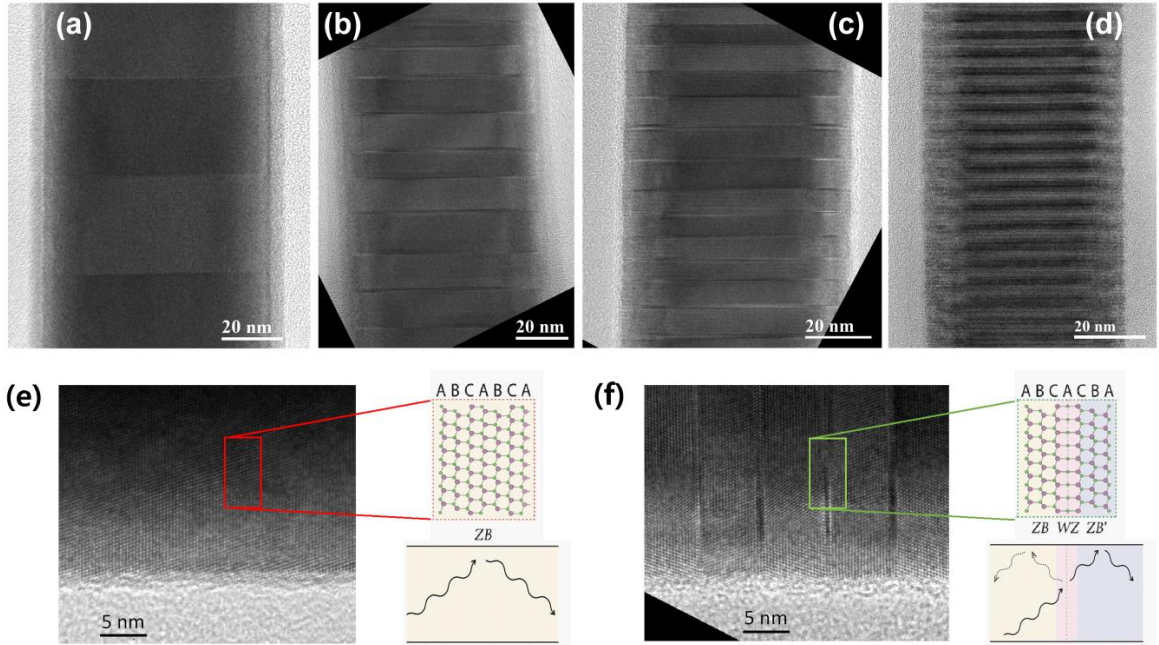

**Figure 1**. HRTEM images of a portion of the InAsSb NWs of samples A (a), B (b), C (c), D (d), with Sb fraction respectively 5.8%, 2,1%, 1.6% and 4.8%. The diameter for the gold catalyzer nanoparticle is 40 nm for samples A, B, C and 20 nm for sample D. The different twin periodicity is clearly seen thanks to the image contrast. Magnified HRTEM image and ball-stick representation of a pure ZB (e) and a TSL (f) InAsSb NW (sample C). The ball-and-stick representations highlight the difference between the crystal stacking order in the ZB phase (ABCABCA) and in the TSL (ABCACBA). In the lower right corner of each panel there is a schematic depiction of phonon scattering in the two different crystal structures: only boundary scattering in the pure ZB NW, with additional scattering at the twin planes in the TSL NW.

| Sample | Au NP diameter (nm) | TBAs line pressure (Torr) | TDMASb line pressure (Torr) | NW average diameter (nm) | $InAs_{(1-x)}Sb_{(x)}$ composition (Sb fraction: x) | Crystal structure (twin periodicity) |
|---|---|---|---|---|---|---|
| A | 40 | 1.4 | 0.3 | 80 ± 10 | 0.058 | Twinned (30 nm) |
| B | 40 | 1.6 | 0.2 | 80 ± 10 | 0.021 | Twinned (12 nm) |
| C | 40 | 1.6 | 0.1 | 70 ± 10 | 0.016 | Twinned (8 nm) |
| D | 20 | 1.5 | 0.2 | 60 ± 10 | 0.048 | Twinned (3.5 nm) |
| E | 20 | 1.5 | 0.4 | 70 ± 10 | 0.072 | Pure ZB |
| F | 60 | 3 | 0 | 70 ± 10 | 0 | WZ |

**Table 1**: Growth parameters and resulting compositional, morphological and structural properties of the NWs measured in the present work.

**Thermal transport properties**

Direct measurements of the thermal transport properties of TSL InAsSb NWs were performed focusing on two different samples among the different grow batches, namely samples C and D, featuring twin periodicity of 8 nm and 3.5 nm, respectively. The performances of these NWs have been compared to those of pure ZB InAsSb NWs (sample E) and those of WZ InAs NWs (sample F). Single NW devices were realized implementing four-electrode architectures in suspended nanowires, specifically designed to perform thermal conductivity measurements via the 3ω method [38-40], an electrical technique exploiting self-heating and non-linear effects for the measurement of the thermal conductivity in high aspect-ratio nanostructures. These devices implement a nanowire suspended at approximately 250 nm from the substrate, as visible in the scanning electron micrograph reported in **Figure 2(a)**. This allows to decouple the nanostructure from the substrate and avoid any heat loss that may detrimentally affect the thermal conductivity measurement. The $3\omega$ method resorts on the measurement of the third harmonic of the voltage drop, $V_{3\omega}$, in the semiconducting nanowire fed with an AC current at frequency omega, $I_\omega$. Figure 2(b-c-d) reports three examples of measured $V_{3\omega}$ signals as function of $I_\omega$ for pure ZB InAsSb NWs (sample E), TSL InAsSb NWs with 3.5 nm (sample D), and TSL InAsSb NWs with 8 nm period (sample C). Together with the cubic fit based on the simplified analytical model [40], we resorted to finite element modeling to exactly solve the temperature and voltage profile in the nanostructure and achieve a precise fit to our experimental data (Supporting Information SI.1). Figures 2(e) and (f) report a comparison between the fit outcomes, revealing the limitations of the purely analytical model which are overcome by employing finite element analysis.

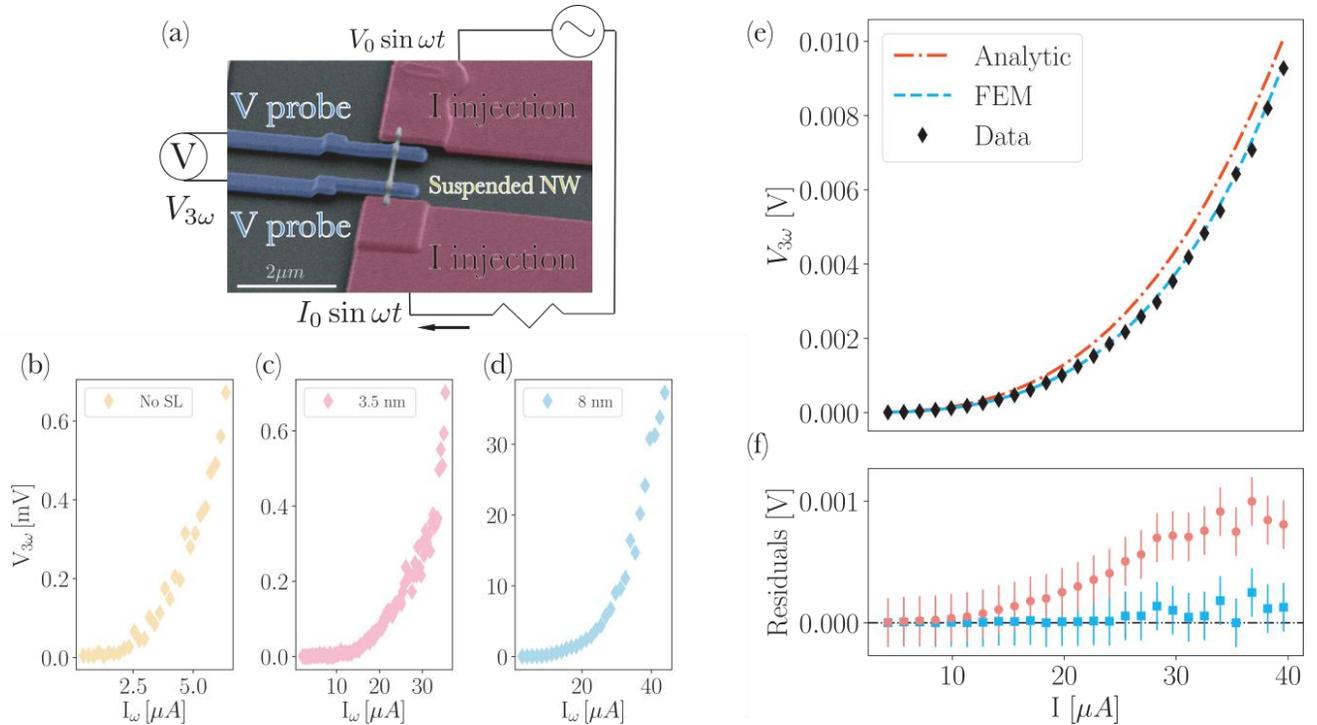

**Figure 2.** (a) SEM image and circuital scheme of a typical NW-based device for thermal conductivity measurements with the $3\omega$ method. In false colors are highlighted the suspended NW (yellow), the injection electrodes (red) which also act as a thermal anchor, and the pick-up electrodes (blue) for a 4-wire measurement of $V_{3\omega}$ with a Lock-In amplifier. Experimental $V_{3\omega} - I_\omega$ curves measured for pure ZB InAsSb NWs (b), TSL InAsSb NWs with 3.5 nm period (c) and with 8 nm diameter (d). (e) An example of a measured $V_{3\omega}$ vs $I_\omega$ curve and the relative fit performed via finite elements analysis compared with the fit performed by employing the analytical formula. (f) Fit residuals for the analytical model and the fit resulting from the numerical calculations. For higher values of $I$, the analytical model deviates from the experimental data, differently from the results of the finite element modeling which show negligible deviation.

The resulting thermal conductivities are reported in **Figure 3**, showing strong evidence of marked reduction of $\kappa$ in all the InAsSb NWs and particularly in TSL InAsSb NWs. Seven independent campaigns of measurements were carried out and the results correspond to the different groups of experimental points reported in Figure 3(a). The thermal conductivity for the WZ InAs NWs (average diameter 70 nm) is found to be $14 \pm 6$ W/mK and is consistent with values reported in literature [38]. A decrease of $\kappa$ is observed in the pure ZB InAsSb NWs, with measured values ranging from $6.3 \pm 2.9$ W/mK to $5.4 \pm 2.8$ W/mK for sample E. This is consistent with the effect on thermal transport of the introduction of Sb in other III-V semiconductor alloys, as reported elsewhere [41] and predicted by the Klemens-Drabble theory, and ultimately stems from the heavier mass of Sb atoms with respect to As atoms. A further reduction is then observed in the InAsSb TSL NWs. The InAsSb NWs with 8 nm twinning period (sample C) exhibit a thermal conductivity ranging from $3.4 \pm 1.4$ W/mK to $2.8 \pm 1.3$ W/mK. The lowest $\kappa$ has been measured for TSL InAsSb NWs with twinning period of 3.5nm (sample D), with an average value of $1.2 \pm 0.6$ W/mK.

Overall, we ascribe the observed experimental results to the enhancement of phonon scattering caused by the incorporation of Sb and by the occurrence of twin planes. In fact, on the one hand the observed reductions of $\kappa$ in ZB InAsSb NWs with respect to WZ InAs NWs cannot be attributed to the change in crystal structure form WZ to ZB. Indeed, the opposite effect – increase of thermal conductivity - has been reported for pure ZB InAs NWs with respect to pure WZ ones [42]. On the other hand, the observed reduction in $\kappa$ could in principle be ascribable to the slight differences between the diameters of the measured NWs, according to the functional dependence expressed by the Ziman's formula [6] which describes how the $\kappa$ of the nanowire deviates from the $\kappa$ of the bulk material as an effect of its diameter:

$$\kappa_{\text{NW}} = \kappa_{\text{Bulk}}\left(1 - e^{-D/\Lambda_i}\right) \qquad (1)$$

where D is the NW diameter and $\Lambda_i$ the phonon mean free path due to internal scattering, which is equal to 250 nm in WZ InAs [6]

To rule out this scenario and take into account the role of the NW diameter, we have defined a normalized $\kappa$, $\kappa_{norm} = \kappa_{NW}/\kappa_{InAs}$, where $\kappa_{InAs}$ is the thermal conductivity of an InAs nanowire with a 70 nm diameter and $\kappa_{NW}$ is the value of the thermal conductivity measured in our InAsSb samples. The dimensionless parameter $\kappa_{norm}$ can be regarded as an estimate of the effective reduction in thermal transport efficiency occurring in our nanostructures. The values of $\kappa_{norm}$ obtained in our samples are reported in Figure 3(b). Starting from $\kappa_{norm}$=1 by definition for sample F (70 nm diameter WZ InAs NWs), we observe a decrease to $\kappa_{norm} = 0.55 \pm 0.19$ in pure ZB InAsSb NW (sample E), and a further decrease to $0.14 \pm 0.06$ and $0.11 \pm 0.04$ for the TSL InAsSb NWs with 8 nm (sample) and 3.5 nm (sample D) twinning period, respectively. In other words, this indicates a nine-fold reduction in thermal conductivity for a TSL InAsSb NW with 3.5 nm twinning period with respect to a WZ InAs NW with the same diameter.

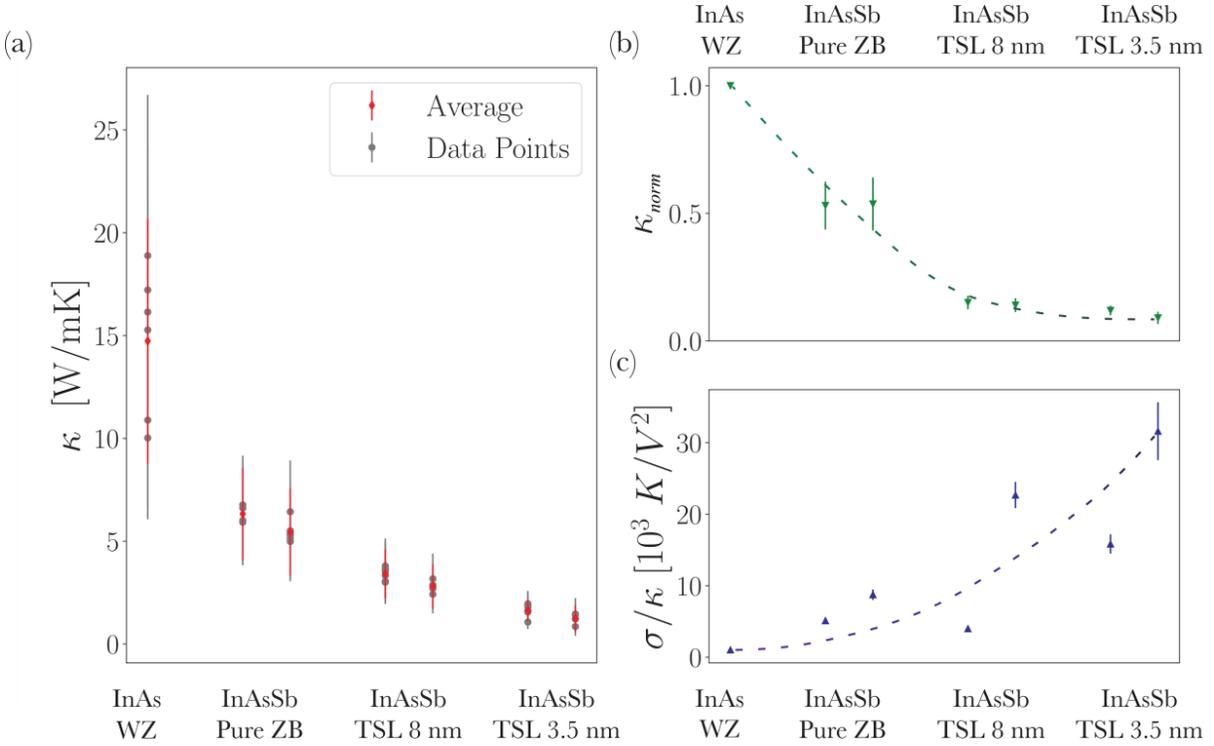

**Figure 3**: (a) Measured thermal conductivity $\kappa$ for the NWs with different crystal structure. Each group of experimental points correspond to an independent campaign of measurements. (b) Thermal conductivity normalized with respect to the value of a 70 nm diameter WZ InAs NW, accounting for the different diameters of the measured NW-based devices. (c) Measured ratio of electrical and thermal conductivity for the different devices, indicating a strong increase in the ratio $\sigma/\kappa$ which appears directly in the thermoelectric figure of merit. Dashed lines in panels (b) and (c) are guides to the eye.

**Electrical and Thermoelectric transport properties**

Together with drastically reduced thermal conductivity, the peculiar polytypic InAsSb heterostructure studied in this work also displays excellent electrical transport properties, equivalent to those of their pure ZB and WZ counterparts. This was ascertained by realizing conventional NW-based field effect transistors (FETs) with the nanostructure laying on a substrate. Here, the nanowires are drop-casted onto a SiO2/Si++ substrate to perform field effect electrical measurement in standard back-gate configuration (Supporting Information SI.2). We fabricated several NW-based FETs starting from sample D (TSL with 3.5 nm period) and sample E (pure ZB) and we measured their electrical transport properties, including the electrical conductivity, σ. This allows us to extract the ratio $\sigma/\kappa$, a key parameter in thermoelectric applications which highlights the impact of the reduced thermal conductivity observed in our samples. Figure3(c) reports the $\sigma/\kappa$ ratio for every measured NW sample, revealing an increase of $\sigma/\kappa$ by one order of magnitude for the best performing TSL InAsSb NWs with respect to the WZ InAs NWs

Starting from the fabricated NW FETs we also measured, at room temperature, the current-back gate voltage characteristic curves, $I_{DS}(V_{GS})$. For each device, from the slope of the linear region of the $I_{DS}(V_{GS})$ curve we obtained the transconductance $m_{lr} = dI_{DS}/dV_{GS}$ that allow to extract the electron mobility, $\mu$ [43] (supporting Information SI.3). **Figure 4.a** reports the electrical mobility measured in all the FETs fabricated with NWs from sample E (ZB InAsSb NWs, blue dots) and sample D (TSL InAsSb NWs with 3.5 nm twinning period, red dots), as function of the electrical conductivity. The results clearly indicate the similarity in electrical transport between pure ZB and TSL InAsSb NWs, with average electron mobility for both materials of approximately 1500 $cm^2/Vs$, also in agreement with the values reported in literature for nominally undoped WZ InAs NWs of similar diameter at room temperature [16, 44].

After $\mu$ is known, we derive the carrier concentration $n = \sigma/e\mu$ and use it to extract the Seebeck coefficient $S_e$ according to the Mott's formula for a degenerate semiconductor [36]:

$$S_e = \left(\frac{\pi^2}{3}\right)^{\frac{2}{3}} \left(\frac{m^* k_B^2 T}{\hbar^2 e}\right) n^{-\frac{2}{3}} \qquad (2).$$

As no significant differences are expected in effective mass for the ZB or TSL InAsSb (with low Sb content) and InAs [27, 23], in expression (2) we can safely assume $m^*$ of InAs. Figure 4(b) displays the extracted Seebeck coefficient as function of σ for pure ZB InAsSb NWs (sample E) and TSL InAsSb NWs with twinning period 3.5 nm (sample D), together with data reported in literature for WZ InAs NWs [16]: also in this case, as for the electron mobility, the similarity between the different nanostructures is quite evident. The spread observed in the measured values of electron mobility and density and of the Seebeck coefficient can be ascribed to the impact of different nanowire diameters (20 nm spread for both twinned and untwinned NW batches) and to fabrication-related factors, e.g., slightly device-dependent contact resistances.

Finally, combining the measured thermal and electrical conductivity and the Seebeck coefficient, we estimate the thermoelectric figure of merit, ZT, of the measured nanostructures, as reported in Figure 4(c). While typical values reported for WZ InAs NWs are in the range of 0.01-0.02 [16, 17], we observe a first improvement for pure ZB InAsSb NWs, with a maximum value of $ZT = 0.042 \pm 0.003$, and a much stronger improvement for TSL InAsSb NWs, with highest reported $ZT = 0.24 \pm 0.07$ in III-V NWs, that is, a net increase in ZT up to an order of magnitude with respect to WZ InAs NWs with the same diameter.

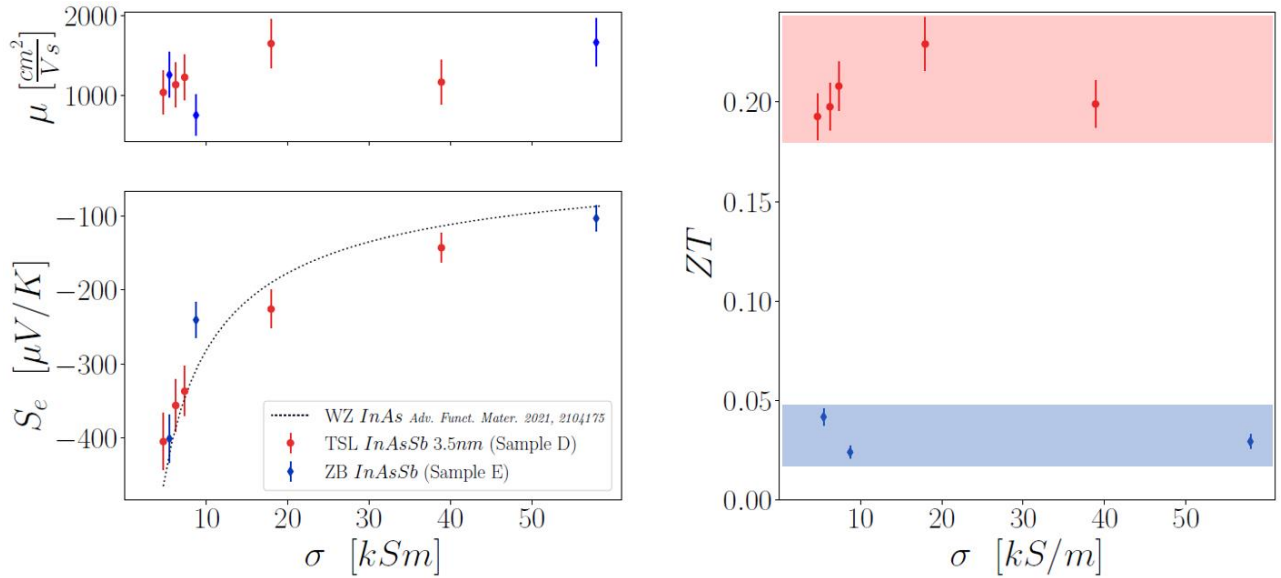

**Figure 4.** (a) Measured electron mobility of the NW-based FETs. As in all other panels, pure ZB InAsSb NWs are shown in blue while red indicates TSL InAsSb (3.5 nm). All the values are perfectly consistent, albeit slightly lower, with respect to electron mobility reported for pure WZ InAs NWs at room temperature [16, 17, 45]. (b) Seebeck coefficient extracted from the transconductance curves via Mott's formula. The dotted line reports the Seebeck coefficient extrapolated from the results presented in literature for WZ InAs NWs. (c) Thermoelectric figure of merit ZT for pure ZB InAsSb NWs (blue dots) and twin superlattice InAsSb NWs with 3.5 nm periodicity (red dots).

**CONCLUSIONS**

We have unveiled experimentally for the first time the correlation between the controlled formation of periodical crystal defects (twins) in NWs and their thermal transport properties, demonstrating a decrease up to one order of magnitude in the thermal conductivity of TSL InAsSb NWs with respect to their pure ZB counterparts. We have reported TSL InAsSb NWs with average diameter of 70nm exhibiting thermal conductivity as low as 1.2 W/mK, that is, an order of magnitude lower than WZ InAs NWs with same diameter. This unprecedented decrease of thermal conductivity due to the occurrence of the twin superlattice was shown to coexist with unaltered electrical conductivity with respect to the InAs NWs, leading to a 10-fold increase in the ratio $\sigma/\kappa$. Besides, almost identical electron mobility and Seebeck coefficient has been extracted for pure ZB and TSL InAsSb NWs, with values in agreement with WZ InAs NWs with similar diameters at room temperature. The giant reduction of k together with unaltered $\sigma$, $\mu$ and $S$ overall yield to a thermoelectric figure of merit for TSL InAsSb NWs one order of magnitude higher with respect to WZ InAs NWs, with the highest reported value of $ZT = 0.24 \pm 0.07$ in III-V NWs. The generality of the twinning process, together with the possibility to engineer TSL NWs with different III-V and IV semiconductors, increase the impact of the present results making them a cornerstone for the development of a new generation of high-performance nanostructures for thermoelectric energy conversion.

**METHODS**
**NWs growth**. InAsSb NWs were grown on InAs(111)B substrates by Au-assisted Chemical Beam Epitaxy (CBE) in a Riber Compact-21 System. The system employs pressure control of the metalorganic (MO) precursors in the gas lines to vary the precursor flux on the sample during growth. The precursors used for the NW growth were trimethylindium (TMIn), tertiarybutylarsine (TBAs), and tert-dimethylaminoantimony (TDMASb). Au nanoparticles (NPs) were obtained by drop-casting of colloidal solution onto the bare substrates. Colloids of 20 and 40 nm diameter were used. After degassing the substrates at 300°C, they were mounted into the CBE chamber and short InAs NW stems were grown using 0.6 Torr and 1.6 Torr of TMIn and TBAs line pressures respectively, at the substrate temperature of 400 ± 5°C. Then, to grow the InAsSb segments, the TMIn line pressure was changed to 0.9 Torr while TBAs and TDMASb line pressures were set to the values reported in Table 1, and the growth temperature was increased to 440°C within 5 min. The InAsSb growth proceeded at this temperature for further 50 minutes. At the end of the growth the TMIn flux was stopped, while the samples were cooled down under both TDMASb and TBAs flow. In Table 1 we summarize the growth parameters, the measured chemical composition and the crystal structure of the NWs.

**NW morphology & structure characterization.** The morphology of the NWs was characterized by means of scanning electron microscope (SEM) using a Zeiss Merlin field-emission microscope at an accelerating voltage of 5 keV. Chrystal structure and chemical composition of the InAsSb segment were analyzed by transmission electron microscopy (TEM) in a JEM-2200FS, equipped with in-column Ω filter and Oxford X-ray energy dispersive spectrometer (EDX), operated at 200 keV. Imaging was performed in high resolution (HR) TEM mode combined with zero-loss energy filtering and by high angle angular dark-field in scanning mode (HAADF-STEM) for the EDX measurements.

**Device fabrication**.
After the growth, the NWs were mechanically detached from the growth substrate by means of sonication and dispersed in isopropyl alcohol (IPA). The adopted fabrication protocol was different for the two investigated device architectures. For standard NW-based FETs, NWs were directly transferred to a pre-patterned $p^{++}Si/SiO_2$ substrate (0.5 mm Si, 280 nm $SiO_2$) by drop-casting a droplet of IPA/NWs solution onto the substrate. In the case of suspended nanowire-based devices for thermal conductivity measurement, a PMMA sacrificial layer was preliminarily spun onto the substrate, and the NWs were drop-casted on top of it. For both the architectures, contact electrodes were then patterned using standard e-beam techniques. After the development and prior to metal evaporation (Ti/Au, 10/100 nm), the NW contact areas were passivated using an ammonium polysulfide $(NH_4)_2S$-based solution to promote the formation of low-resistance ohmic contacts [45]. In the case of suspended NW-based devices, after liftoff, the sacrificial layer underneath the contacts was removed by means of conventional oxygen plasma dry etching. See also Supporting Information SI.2-3.

**Finite Element Modeling.**

In its simplest formulation, the $3\omega$ method relates the third harmonic of the voltage drop across a NW to the injection current as [40] $V_{3\omega} = \left(\frac{4LRR'}{\pi^4 S\kappa}\right)I^3$. This result, however, is an approximation that results from truncating the Fourier series expressing $V_{3\omega}$, and it can be shown to hold only in nanostructures of high thermal conductivity. Considering the giant reduction of $\kappa$ observed in TSL NWs, a simple application of the cubic model to our dataset would lead to an underestimate of the NW thermal conductivity.

To account for that, the data analysis has been performed via a Finite Element Model, which is able to account for the non-linearities in the transport equations caused by the drastically reduced $\kappa$ [16, 40, 38], as evident from Figure 2. The NWs were modeled as hexagonal prisms, with height and base were measured form the SEM images and TEM images respectively.

Our FEM model solves the coupled physics problem of heat equation and the local version of Ohm's law to account for Joule's heating and the consequent local change in electrical resistivity. The current, voltage and temperature profiles are then computed via the FEniCS package [46], allowing us to extract $V_{3\omega}(I)$ for each value of injection current experimentally used. This method allows us to perform a least-squares fit on the experimental data with the thermal conductivity as the only free parameter, from which we obtained the values of $\kappa$ discussed in the main text. See also Supporting Information SI.1.

20 1.

[43] D. Prete, V. Demontis, V. Zannier, M. J. Rodriguez-Douton, L. Guazzelli, F. Beltram, L. Sorba, F. Rossella, Nanotechnology 2021, 32, 14 145204.

[44] A. C. Ford, H. Johnny C., Y. L. Chueh, Y. C. Tseng, Z. Fan, J. Guo, J. Bokor, A. Javey, Nano Letters 2009, 9, 1 360.

[45] D. Suyatin, C. Thelander, M. Björk, I. Maximov, L.Samuelson, Nanotechnology 2007, 18, 105307.

[46] A. Logg, G.N. Wells, ACM Transactions on Mathematical Software, 2010, 37,2.

# Supporting Information

## SI.1. Finite Element Analysis

The equations to be solved to compute the electrical potential and temperature profiles of the system under analysis are the following thermoelectric equations:

$$\begin{cases} \boldsymbol{j}(\boldsymbol{x},t) = \sigma(\boldsymbol{x},t)\boldsymbol{E}(\boldsymbol{x},t) \\ \boldsymbol{\Phi}(\boldsymbol{x},t) = \kappa \nabla T(\boldsymbol{x},t) \end{cases}$$

together with the heat equation, which can be written as follows:

$$\kappa \nabla^2 T(\boldsymbol{x},t) = \boldsymbol{j}(\boldsymbol{x},t) \cdot \boldsymbol{E}(\boldsymbol{x},t)$$

where $\boldsymbol{j}$ is the electrical current density in the nanowire, $\sigma$ is the electrical conductivity, $\boldsymbol{E}$ is the electric field, $\boldsymbol{\Phi}$ is the heat flux in the nanowire, $\kappa$ is the thermal conductivity, $T$ is the temperature.

The boundary conditions imposed to solve the problem are:

$$\begin{cases} T(0,t) = T_0 \\ T(L,t) = T_0 \end{cases}$$

where L is the nanowire length and $T_0$ is chosen to be room temperature. Furthermore:

$$\begin{cases} V(0,t) = 0 \\ \oint \boldsymbol{j}(\boldsymbol{x},t) \cdot d\boldsymbol{S} = 0 \end{cases}$$

where the second equation is needed to ensure that no charge is accumulated in the nanowire.

The finite element analysis is implemented with the FEniCS python packages [1-2], by exploiting the GMSH software in order to design the domain and produce the mesh used as an input for the python-implemented solver.

The computation of $V_{3\omega}$, as well as the first harmonics, to be compared with experimental data are achieved by placing two voltage probes at each end of the NW, 50 nm away from the prism bases where the boundary conditions were placed, mirroring the actual experimental conditions reported in the main text. Since time appears merely as a parameter modulating I(t), the voltage drop was computed 64 times each $\frac{1}{64}\frac{2\pi}{\omega}$ to span a full period of the current oscillation, and then the Fast Fourier Transform [3-4] of the signal was computed and normalized by a factor 1/64. After this, the second and fourth entry in the obtained array are, respectively, $V_\omega$ and $V_{3\omega}$.

To compute the value for the thermal conductivity, the fit of the experimental curves of $V_{3\omega}$ with respect to $I_0$ was performed with the FEM simulations through the built-in scipy [5] implementation of the Levenberg-Marquardt algorithm for least-square fits. The fitting parameter in the case of the FEM simulations was $\kappa$ directly, as *curve_fit* was called on a function taking $\kappa$ and $I_0$ as parameters and returning the respective $V_{3\omega}$ obtained as described earlier. It is thus important to stress that $\kappa$ was the only free parameter in the simulation. To speed up the fit convergence, as computing $V_{3\omega}(I_0)$ for each data point is quite computationally heavy, the initial value for the thermal conductivity in the FEM fit was set as the value obtained though the cubic fit according the analytical approximated relation: $V_{3\omega} = \frac{4R_0 R' L}{\pi^4 S \kappa} I_0^3 = A I_0^3 \rightarrow \kappa = \frac{4R_0 R' L}{\pi^4 SA}$, where $R_0, R', L$ and $S$ and the resistance of the nanowire at room temperature, the temperature coefficient of the resistance, the length and area of the nanowire, respectively.

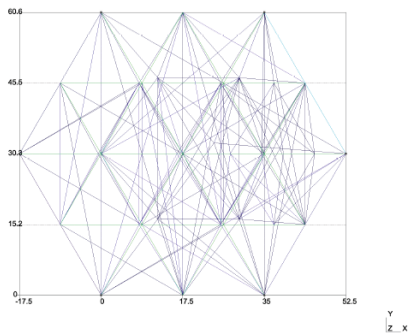

(a)

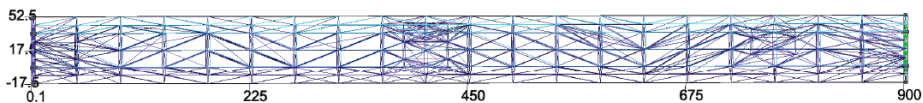

(b)

Figure S1: Mesh of the domain employed for the finite element analysis calculations. (a) front cross section and (b) lateral cross section of the geometry.

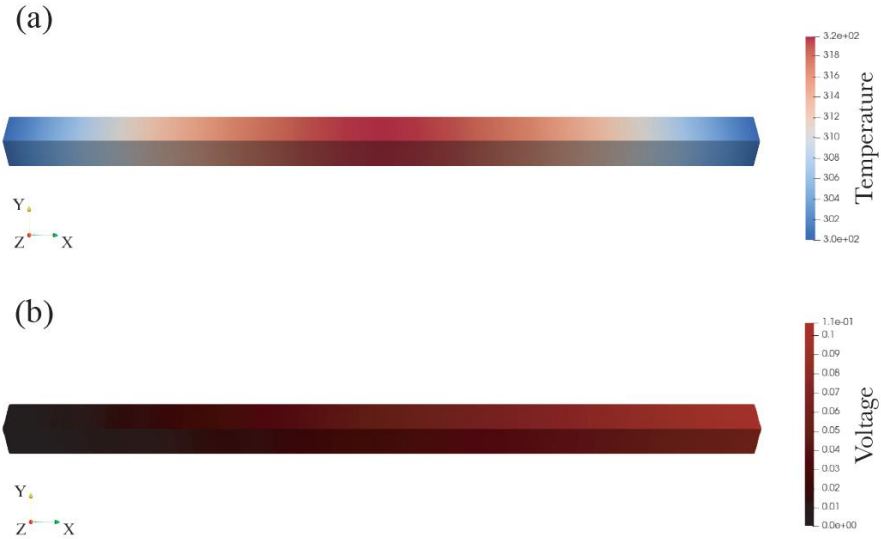

Figure S2: Temperature and voltage profiles resulting from a typical calculation, for a constant value of injected AC current with amplitude 10 $\mu A$. (a) Computed temperature profile, showing a maximum in the centre of the nanowire. (b) Computed voltage profile, showing an overall voltage drop between the current injection face and the opposite side of the nanostructure.

## SI.2. Nanowire field effect transistors realization

Conventional dropcasted nanowire-based field effect transistors were fabricated to perform electrical characterization of the nanostructures. The fabrication substrates (a Si++/SiO$_2$ bilayer) were pre-patterned via UV lithography to define the contact pads to perform electrical transport measurements. The substrate itself can be employed as a back-gate to perform field effect measurements and extract the relevant parameters for the electrical transport characterization.

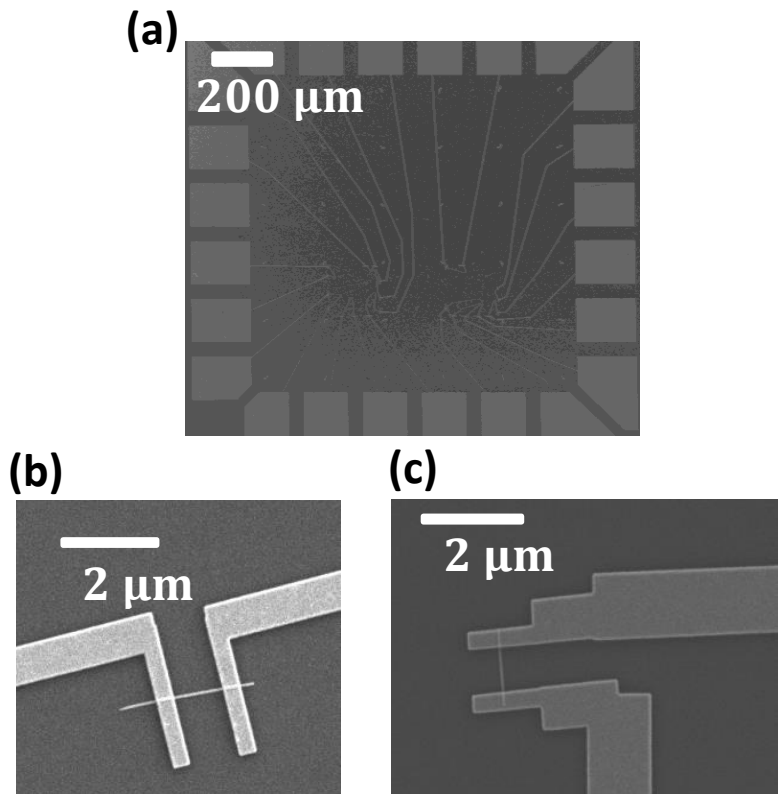

Figure S3: Prototypical dropcasted devices for electrical transport characterization. (a) Devices are fabricated via electron beam lithography on pre-patterned chips with defined gold pads for electrical measurements. (b)-(c) Zoom-in of the scanning electron micrograph of panel (a) showing in detail the structure of the nanowire-based field effect transistors employed for the electrical transport measurements

# SI.3. Electrical transport characterization

In order to compare the electron concentration and mobility for the nanowire samples, we probe the electrical transport properties with field effect measurements exploiting Si$^{++}$/SiO$_2$ back-gating. In this configuration, we can obtain an estimate for the electron mobility by referring to the known relations for MOSFETs [6]:

$$\mu = \frac{L}{W}\frac{1}{V_{DS}C_A}m_{lr}, \quad (2)$$

where L, W are the channel length and width respectively, $V_{DS}$ is the applied bias voltage, $C_A$ is the gate capacitance per unit area and $m_{lr}$ is the slope of the transconductance in the linear regime, i.e.

$$m_{lr} = \frac{dI_{DS}}{dV_{BG}}\bigg|_{linear}. \quad (3)$$

We evaluate the total capacitance of the nanowire/back-gate system approximating it as a cylinder with length L and radius r on an infinite plane, with the dielectric separating the cylinder and the charged plane where the voltage is applied having a thickness $d_{SiO_2}$. In this configuration, the total capacitance is:

$$C = \frac{\pi \epsilon_{SiO_2} L}{\operatorname{arccosh}\left(\frac{d_{SiO_2}+r}{r}\right)}, \quad (4)$$

where $\epsilon_{SiO_2} = 3.9\epsilon_0$ is the electrical permittivity of SiO$_2$. By substituting in (2) the capacitance per unit area with the total capacitance reported in (4) and noting that $C = C_A L W$, it follows that:

$$\mu = \frac{L^2}{CV_{DS}}m_{lr} \quad (5)$$

Finally, the carrier concentration can be estimated by noting that in the diffusive limit the Drude relation holds:

$$\sigma = ne\mu \rightarrow n = \frac{\sigma}{e\mu} \quad (6)$$

where $\sigma$ is the electrical conductance, expressed in units of S/m.

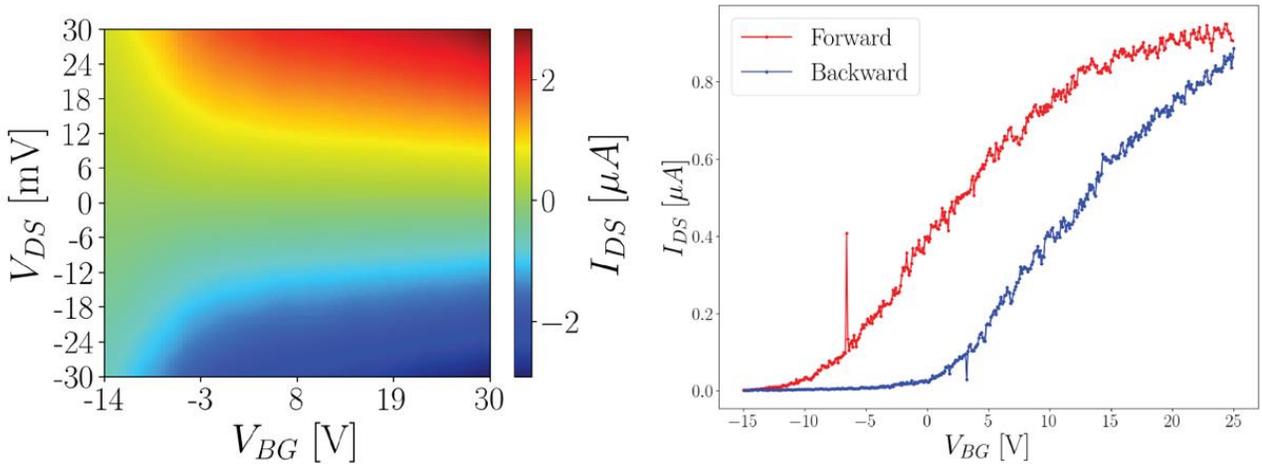

Figure S4: Field effect transistor operation for a device fabricated with pure zincblende InAsSb nanowires.

| $R\ [k\Omega]$ | $\mu\ \left[\frac{cm^2}{Vs}\right]$ | $n\ [10^{17} cm^{-3}]$ |
|---|---|---|
| 6 | 1644 | 22 |
| 40 | 881 | 6.1 |
| 22 | 1265 | 5.7 |

Table S1: Extracted values for the electrical resistance, mobility and carrier concentration for ZB InAsSb nanowires

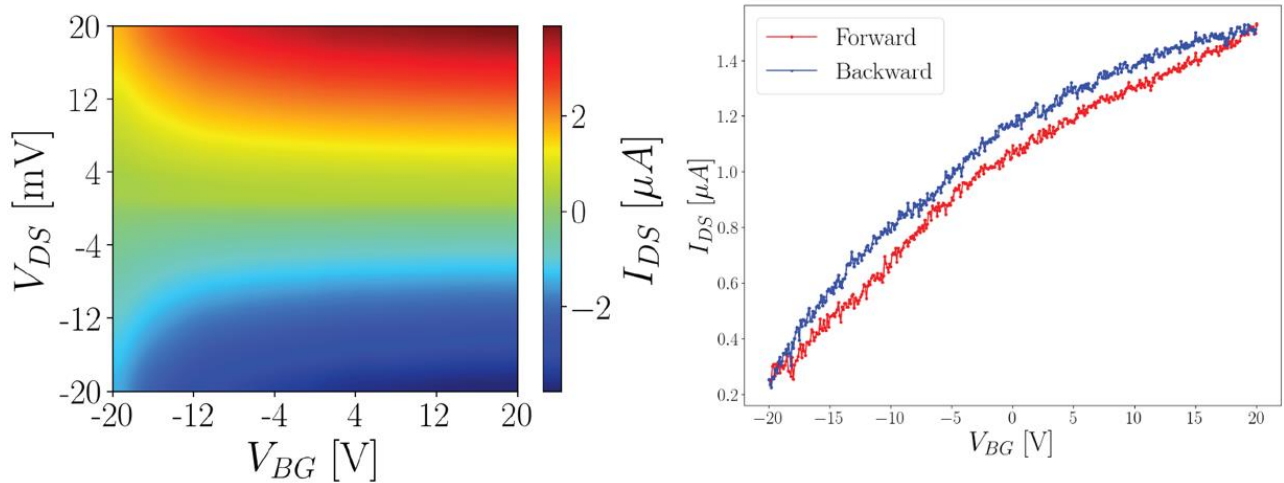

Figure S5: Field effect transistor operation for a device fabricated with twinned superlattice InAsSb nanowires with twinning period of 3.5 nm.

| $R\ [k\Omega]$ | $\mu\ \left[\dfrac{cm^2}{Vs}\right]$ | $n\ [10^{17} cm^{-3}]$ |
|---|---|---|
| **14** | 850 | 18 |
| **9.5** | 1800 | 13 |
| **8.5** | 23 | 5.7 |
| **13** | 717 | 23 |

Table S2: Extracted values for the electrical resistance, mobility and carrier concentration for twinned superlattice InAsSb nanowires (twinning period 3.5 nm)